# Recent advances in structural and dynamical properties of simplified industrial nanocomposites

A.-C. Genix, G. P. Baeza[ϒ], and J. Oberdisse*

*Laboratoire Charles Coulomb (L2C), UMR 5221 CNRS, Université de Montpellier, F-34095 Montpellier, France*

* Author for correspondence: julian.oberdisse@umontpellier.fr

26[th] of August 2016

- invited feature article -

## Abstract

A large body of experimental work on the microstructure and dynamics of simplified industrial nanocomposites made of disordered silica filler in a styrene-butadiene matrix by solid-phase mixing is regrouped and critically discussed in this feature article. Recent results encompass systems with varying polymer mass, grafting functionality, and filler content. They have been obtained by simulation-based structural modelling of nanoparticle aggregate size and mass deduced from small-angle scattering and transmission electron microscopy. Our model has been validated by independent swelling experiments. Comparison of structurally-close nanocomposites of widely different chain mass led to the identification of a unique structure-determining parameter, the grafting density, as well as to a unified picture of aggregate formation mechanisms in complex nanocomposites during mixing. In addition, low-field proton NMR allowed for the characterization of dynamically slowed-down ('glassy') polymer layers, which were shown not to dominate the rheological response, unlike the structural contribution. Finally, broadband dielectric spectroscopy was used in an innovative manner to identify filler percolation – also identified by rheology – via dynamics along filler surfaces.

[ϒ] Present address: INSA Lyon, Lab MATEIS, CNRS UMR 5510, F-69621 Villeurbanne, France

# I. Introduction

Polymer nanocomposites are colloidal systems of in general inorganic nanoparticles (NPs) embedded in a soft polymeric matrix [1-3]. While the chemical nature – quantity, type and surface modification of NPs, properties of the polymer – are the main parameters defining an experimental system, the way they are mixed and possibly linked, and the dispersion that is finally obtained is crucial for the properties of such nanocomposite samples [4-13]. Examples of dispersions range from individually dispersed NPs [14, 15], to aggregates [16-19], and percolating networks [20], with obvious consequences of such structures on dynamical properties. Percolation, e.g., introduces a rigid scaffold pervading the entire sample, thereby creating connected paths for transmission of stresses as well as conduction of ions. Such effects are measurable by either rheology – the reinforcement effect [17, 21, 22] – or by dielectric methods [23, 24]. Interestingly, one of the major techniques for structural analysis, small-angle scattering [25], is incapable of distinguishing percolating from non-percolating states, and even electron microscopy can only give a hint. This observation brings us immediately to the conclusion that combination of techniques may be the motor of progress in this field.

The sensitivity of macroscopic properties of polymer nanocomposites to structural details motivated a large body of fundamental research on structure [1-3, 5, 18, 26-29]. In the vast majority of cases, work focused on the microstructure (i.e., dispersion) of model colloids, usually spherical and if possible monodisperse nanoparticles in size [18, 30]. While they are much more used in industrial applications due to their optimized properties, industrial nanofillers are not particularly well suited to extract more detail than just some average fractal dimension of aggregates from scattered intensities [21, 31-33], due to their disordered nature. This is why qualitative observations made by TEM are mostly reported in the literature [34, 35]. In spite of this inherent difficulty, several advances may be noted over the past few years. Shinohara et al. have studied both structure and dynamics of nanocomposites with industrial silica in styrene-butadiene (SB) [36]. They have performed SAXS experiments on nanocomposites and presented (to our knowledge) the first *qualitative* interpretation of the multiscale structure based on aggregates and their interaction. Whereas Bouty et al. have explored the effect of coating and coupling agents on the dispersion and mechanical properties [37], a coherent body of research has recently been dedicated to simplified industrial nanocomposites, a term to be defined below [24, 38-43]. This work started with a structural study, as a function of filler content and polymer grafting, allowing the construction of a general model of aggregate formation in such complex systems, as well as the identification of a unique structure-determining parameter. It was then extended to dynamical studies, complementing the structural ones, with a particular focus on immobilized polymer layers [44, 45] and (de-)percolation. The objective of the present feature article is to summarize and put into perspective these results and main conclusions. After presenting the sample preparation, the first part of the study is focused on nanoparticle structure using SAXS and TEM. In a second step, once the static framework of filler dispersion in space clearly established, the extension towards dynamics using dielectric spectroscopy is presented.

## II. Simplified industrial nanocomposite samples

Compared to the complex samples usually studied in the literature, the system investigated here has been designed to reproduce key features of industrial samples, but of simplified composition. Ingredients have thus been limited to the strict minimum, conserving only silica (with a compatibiliser) and graftable polymer chains, all related to tire applications despite the absence of crosslinking.

The polymer matrix of nanocomposites is made of two types of un-crosslinked chains, either graftable or non graftable, with antioxidants. Each chain is a statistical copolymer with styrene (26%w) and butadiene (74%w) units (41% of which are 1-2 and 59% of 1-4), which has been purpose-synthesized by Michelin. Graftable chains bear a single $SiMe_2$-OH function denoted D3. During the mixing process, this functional group may graft the chain on the silica by condensation with the surface silanol. The chain masses, $M_{SB}$, are 40, 80, 140, 190, and 280 kg.mol$^{-1}$, all with polydispersity index below 1.1. Note that, in all cases, we chose to mix graftable and non graftable chains of identical mass. The glass-transition temperature is -35°C (20 K/min heating rate) for the different chain mass. It was not found to vary with the silica content (up to $\Phi_{si}$ = 21.1%v). The silica pellets (Zeosil 1165 MP from Solvay) have a nominal specific surface of 160 m$^2$/g, and the size distribution of the primary nanoparticles obeys a log-normal law in agreement with TEM studies ($R_0$ = 8.55 nm, $\sigma$ = 27%, leading to an average bead volume of $V_{si}$ = 3.6 10$^3$ nm$^3$). A coating agent, octyltriethoxysilane (octeo, 8%w with respect to silica) and a catalyser, diphenyl guanidine (DPG, 1%w with respect to polymer) were also used during the mixing process to improve the filler fragmentation. For consistency, octeo has been added to all samples including those with graftable chains.

Nanocomposites are formulated by stepwise introduction and mixing of the SB chains with silica pellets and octeo/DPG in an internal mixer. Note that particular care was taken to avoid any trace of carbon black, catalysing nanoparticles (ZnO), crosslinking or coupling agents, which may impede interpretation of, e.g., scattering experiments. Effective grafting was checked by bound rubber measurements giving the quantity of non-extractable chains of 75% [39]. The final silica volume fractions are in the range from 8.4%v to 21.1%v. For lower silica contents, inhomogeneous composites were obtained due to a less effective mixing process.

## III. Multi-scale filler structure

In this section, we focus on the filler structure in highly loaded nanocomposites, using the combination of TEM and X-ray scattering. The structure of the silica has been modeled in a step-by-step, multiscale approach, starting with the primary silica nanoparticles as basic building units. [38] These NPs are found to be aggregated in small clusters with the typical radius of about 40 nm. These aggregates are themselves concentrated in large-scale fractal branches (thickness ca. 150 nm, extending over micrometers), where they repel each other via a repulsive interaction potential.

### III.1 Influence of filler content on filler microstructure

The filler content was investigated in a series of samples at fixed matrix composition of 50%D3 (50% reactive chains) with 140 kg.mol$^{-1}$ chains, and different silica volume fractions. [38] In Figure 1a, representative TEM pictures for the large-scale structure of two samples ($\Phi_{si}$ = 8.4 and 21.1%v) are shown.

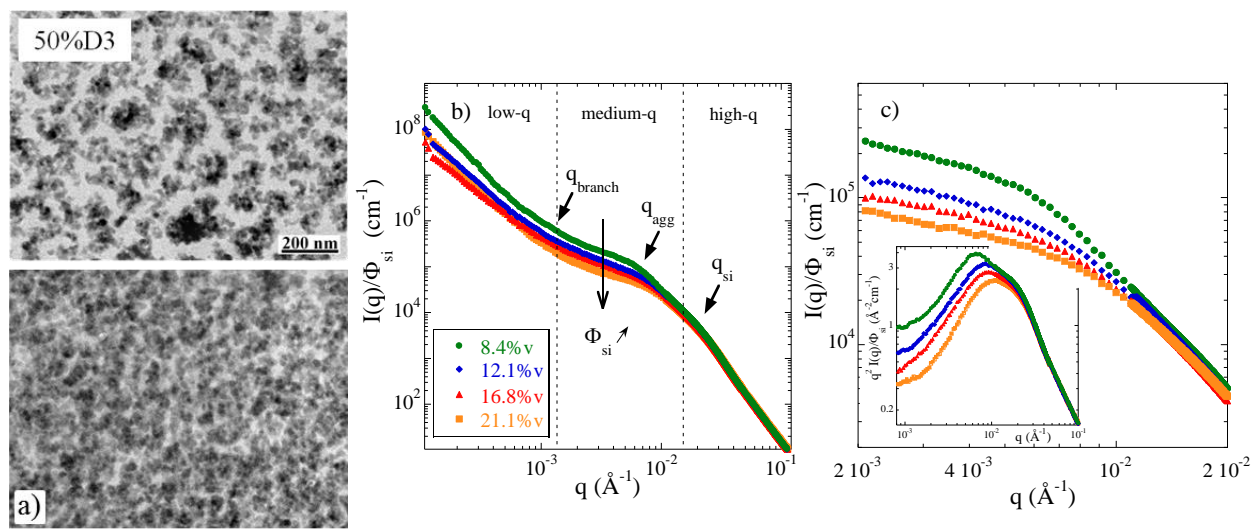

**Figure 1:** Silica structure of nanocomposites (140 kg.mol$^{-1}$, 50%D3). **(a)** TEM-pictures with $\Phi_{si}$ = 8.4%v (top) and 21.1%v (bottom). **(b)** Reduced SAXS scattered intensity for a series of silica fractions in matrix (8.4 – 21.1%v). Arrows indicate the breaks in slope discussed in the text. **(c)** Intermediate-q structures highlighted after subtraction of the low-q power law. Inset: same data in Kratky presentation. Adapted with permission from G. P. Baeza et al. Macromolecules 46 (2013) 317-329. Copyright (2013) American Chemical Society.

The structural features in the 8.4%v-sample can be summarized as follows: a dense branched structure of lateral dimension of about 150 nm is seen, with a significant amount of empty space (pure polymer zones) between the branches. The latter are made of small silica NPs of approximately 10 nm radius aggregated together. A grey-scale analysis of the pictures revealed that the pure polymer fraction is about 41 ± 4% in surface. Note that in the slice, most of the silica NPs are visible individually, leading to a first level of grey, whereas a small number overlap giving a darker grey. From simple geometric considerations, it appeared that in thin enough slices (about 70 nm in the present case), of thickness smaller than the structural length under study in the sample (≈ 150 nm), the surface and volume fractions of matter (branches) coincide. It was thus concluded that approximately 41% of the sample is not occupied by branches. Similarly, the higher volume fraction sample shown in Figure 1a is much denser, with a pure polymer fraction of about 20 ± 4% in surface (and thus also in volume).

Direct imaging methods like TEM have the advantage of intuitive analysis. On the other hand, they convey limited information at high filler concentration where structural features become hardly distinguishable. This is not the case of reciprocal space methods like SAXS, which are highly representative – though more difficult to interpret – as we will see now. The scattering data are shown in Figure 1b for the series in silica fractions. In the reduced representation, $I(q)/\Phi_{si}$, the curves are seen to overlap perfectly above a critical wave vector $q_{si} \approx 0.01$ Å$^{-1}$, corresponding to primary silica NPs in close contact. There is a strong low-q upturn at q-values below $q_{branch} \approx 10^{-3}$ Å$^{-1}$, which has been represented by a power law $Aq^{-d}$, with fractal dimension d = 2.4 ± 0.3. In this representation, the value of the prefactor A decreases with increasing $\Phi_{si}$. We will see below that this is related to the decrease in isothermal compressibility at intermediate q-values.

Figure 1c focuses on the intermediate and high-q ranges, after subtraction of the low-q power law. A slowly varying scattering curve has been obtained below $5.10^{-3}$ Å$^{-1}$ for all silica fractions, and a clear decrease of the scattered intensity is observed for increasing $\Phi_{si}$. A quantitative model for the structures identified in this range has been developed recently [38]. It gives considerably more details on aggregate structure than the traditional power law analysis [46]. The main ingredients are recalled here: In the intermediate-q range, the remaining intensity can be described in terms of polydisperse aggregates interacting inside the branches. Their average radius – ca. 40 nm – has been deduced from a Kratky analysis (inset in Figure 1c), the maximum and the shoulder there corresponding to $q_{agg}$ (resp. $q_{si}$) indicated in Figure 1b. A central piece of this analysis is the description of the inter-aggregate interaction by a simulated structure factor $S_{inter}(q)$ for polydisperse spheres representing aggregates. $S_{inter}(q)$ have been calculated by Monte Carlo simulations assuming excluded volume interactions between aggregates of polydispersity in of 30%. $S_{inter}(q)$ corresponds to the structure factor of an infinite homogeneous sample of aggregates at aggregate volume fraction $\Phi_{agg} = \Phi_{si}/(\kappa\Phi_{fract})$, as aggregates are located within the branches. $\Phi_{fract}$ is the volume fraction of fractal branches determined by TEM. Because of the effect of aggregate compacity and of pure polymer zones, $\Phi_{agg}$ is significantly higher in the branches than $\Phi_{si}$. The point is that the low-q limit of $S_{inter}(q)$, which can be associated with an apparent isothermal compressibility, is lower for more concentrated samples. By continuity, this intensity depression is passed on to the structure factor describing the fractal: the complete scattering curve is thus lowered in the intermediate- and low-q ranges.

Comparison with the experimental intensity at a low-q value ($10^{-3}$ Å$^{-1}$) led to the determination of the average aggregate compacity (assumed identical for all aggregates in the distribution, 31 – 38%), and thus the distribution in aggregation number, which was found to be very large: $\Delta N_{agg}$ is of the same order of magnitude as $\langle N_{agg}\rangle$, i.e. about 45, with only minor changes with increasing silica fraction. This is surprising and in apparent contradiction with practical knowledge on solid-mixing in industry. It is clear from our analysis that interpreting the intensity decrease with $\Phi_{si}$ at intermediate-q (Figure 1c) as a decrease of the aggregate form factor only could lead to the erroneous interpretation of decreasing aggregate mass, instead of additional crowding within the fractal branches for higher silica loadings. Note that the reinforcement effect of these silica structures in the SB-matrix has been characterized by rheology and described with a model based on the aggregate compacity, see section IV.1 [38].

### III.2 Influence of matrix composition (grafting)

After the influence of the filler content, the second important control parameter was the addition of graftable polymer chains being able to covalently bind on the filler surface. Their fraction among all chains was varied from 0 to 100%D3, the remaining chains forming the matrix [39], all at 140 kg.mol$^{-1}$. In Figure 2a, the TEM pictures of nanocomposites containing 8.5%v of silica and different grafting fractions are compared. Pictures for intermediate grafting fractions can be found in the original article.

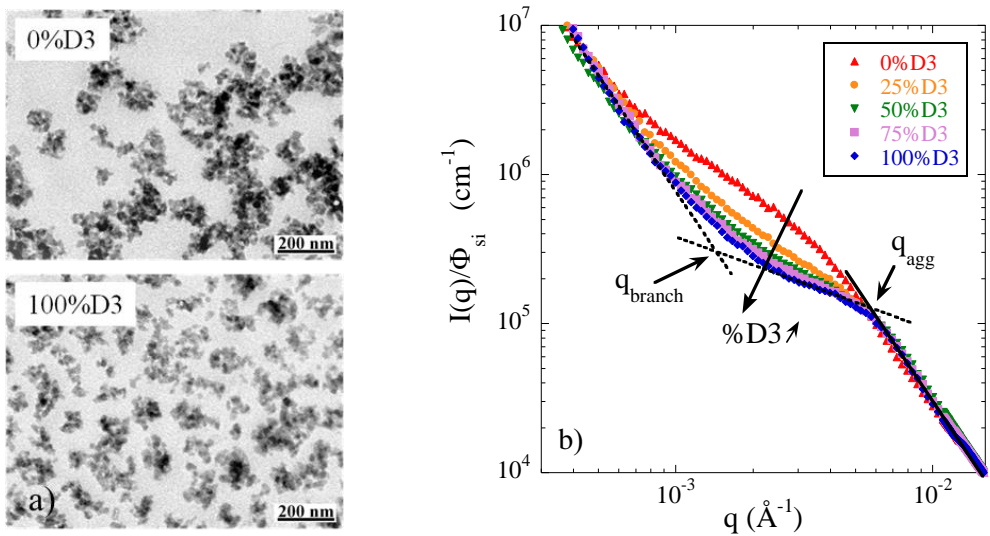

**Figure 2:** Silica structure of 8.5%v-nanocomposites (140 kg.mol$^{-1}$) with different matrix compositions from 0 to 100%D3. **(a)** TEM-pictures. **(b)** Reduced SAXS scattered intensities. Arrows indicate the breaks in slope. Adapted with permission from G. P. Baeza et al. Macromolecules 46 (2013) 6388–6394. Copyright (2013) American Chemical Society.

The large-scale structure of these nanocomposites evolves from a rather heterogeneous spatial distribution at 0%D3 to a much more homogeneous dispersion at 100%D3. Without grafting functions in the matrix, connected polydisperse aggregates forming large irregular branches are observed, leaving space to large zones without silica. More regular branches are found from 50%D3 on (not shown here). When all chains bear grafting end functions, a possible breakup of these fractal branches appears. These branches are well dispersed in space, with "channels" of approximately constant width separating the "islands" of aggregates.

The effect of matrix composition (%D3) on the small-angle scattering results is demonstrated in Figure 2b, at a fixed silica fraction of 8.5%v. Similar data for 16.7%v may be found in the original article. From the position of the low-q crossover, the average thickness of the branches making up a fractal-like structure was found to evolve from 220 to 90 nm for increasing %D3 at 8.5%v silica. This seems to correspond to the decrease of the large channels at 0%D3 to thinner ones at 100%D3 in the TEM pictures. The intensity in the intermediate-q range depends on the D3-fraction: as the grafting fraction increases, the intensity level decreases continuously for both $\Phi_{si}$. One may notice that the decrease becomes clearly less pronounced for D3-fractions above 50%. This is a first indication for a saturation effect, and we will come back later to this point.

In the intermediate-q range, the intensity decrease seems to be similar to the one observed previously when increasing $\Phi_{si}$. However, the situation is different here due to the fixed silica fraction, implying that the aggregates organize differently in space. This effect has been analyzed quantitatively by applying the model of interacting aggregates referred to in section III.1. In absence of grafted chains at $\Phi_{si}$ = 8.5%v, the average aggregate radius is about 50 nm, and the resulting aggregate compacity is very high, close to 55%, giving a high average aggregation number of about 160. With increasing fraction of grafted chains, the compacities were found to fluctuate around 35%, with a roughly constant aggregate radius of 40 nm, and aggregation numbers decreasing to around 60. The same analysis was applied to the 16.7%-nanocomposites, and similar but steeper evolutions were found. These results reflect the saturation in intensity observed in Figure 2b: after a rather abrupt change, aggregate radius, compacity and aggregation number level off to more or less constant values for D3 values above 25%. Such a saturation effect of the structural parameters with the fraction of graftable chains suggests a threshold in grafting density below which the insufficient coverage of the silica surface makes the system prone to structural changes.

### III.3 Influence of chain mass

Due to the chain-mass dependence of the mixing viscosity, it is tempting to believe that the final filler structure is determined by the chain mass. It was thus obvious for us to investigate the impact of this parameter on the structural features, after $\Phi_{si}$ and the grafting fraction. Note that the combination of these three control parameters leads to a single one, the grafting density, which will be discussed in the next sub-section. The chain mass has been varied from 40 to 280 kg.mol$^{-1}$ at fixed matrix composition (50%D3), with the same molecular weight for the two types of chains [40]. The number of chain ends per unit silica surface is thus decreasing with increasing chain mass. In Figure 3, the scattered intensities of nanocomposites of fixed silica fraction are shown for four different chain masses.

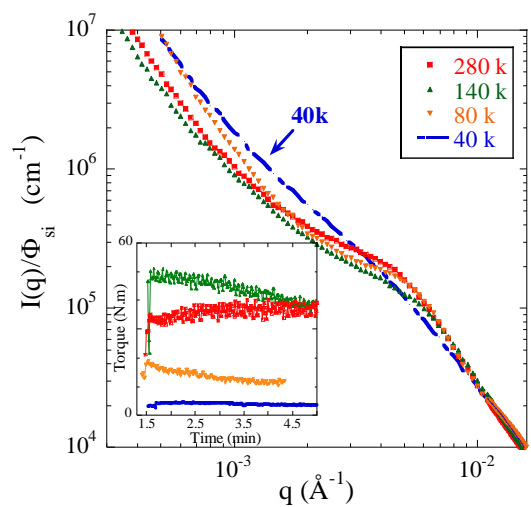

**Figure 3:** Reduced SAXS scattered intensity for a series of nanocomposites (50%D3) with different polymer chain masses at $\Phi_{si} \approx 9.5\%v$. Inset: torque measured during the mixing process of the same samples. Adapted from Ref [40] with permission from the Royal Society of Chemistry.

In Figure 3, a singular sample needs to be discussed first. Unlike all other samples, the 40 kg.mol$^{-1}$ sample is a highly viscous liquid, which induces a very low torque in the mixer (see inset in Figure 3). For comparison, we have measured the structure of the (raw) silica pellet powder by SAXS, which was found very close to the 40k-composite. It can thus be concluded that the viscous nanocomposite sample contains only dispersed original silica powder, without fragmentation of the silica pellets into nano-sized objects, due to an insufficient mixing torque.

The remaining scattering curves in Figure 3 display the same features as all curves discussed previously. The aggregate radii were determined by the Kratky plots, and the intensity level was used to calculate the aggregate compacity. With $\approx 9.5\%v$ of silica, the values obtained for $<R_{agg}>$ were of the order of 45 nm for masses between 80 and 280 kg mol$^{-1}$, and the compacity varied between 30 and 40%, with average aggregation numbers from 50 to 100. At the higher silica fraction (not shown) [40], the aggregate size was found to increase with the chain mass. $N_{agg}$ more than doubled from 40 to 280 kg.mol$^{-1}$, while the compacity stayed again approximately constant in the 35 – 40% range.

### III.4 Towards a single structure-determining parameter

The difficulty with the data presented up to now is that the quantitative characterization has to be carried out for each silica fraction, for each fraction of graftable chains, and for each chain mass. As a next step, a unique structure-determining parameter for these simplified industrial nanocomposites has been proposed [40, 41]. We have seen in the preceding section that higher chain masses at fixed matrix composition lead to bigger aggregates, an effect opposite to higher chain grafting. The latter two dependencies together suggest checking the pertinence of the grafting density, $\rho_{D3}$, as a control parameter

$$\rho_{D3} = \frac{(1-\Phi_{si}) \, \%D3 \, d_{SB} \, N_A \, R_0 \, \exp(2.5\sigma^2)}{3 M_{SB} \, \Phi_{si}} \quad (1)$$

$d_{SB}$ = 0.94 g·cm$^{-3}$ is the density of the polymer, and $N_A$ is the Avogadro number. Eq. (1) is calculated with the average surface and volume deduced from the log-normal NP size distribution ($R_0$, $\sigma$). $\rho_{D3}$ corresponds to the nominal number of grafting functions per unit silica surface. For all available samples, the grafting density has been calculated, and couples (so called twins) identified with $\rho_{D3}$ as close as possible [41]. In some cases, e.g., a doubled mass compensated by doubled grafting, the comparison may be exact.

In Figure 4a, the comparison between the TEM pictures (1-2) for twin samples with $\rho_{D3} = 35\times10^{-3}$ nm$^{-2}$ is shown ($\Phi_{si} \approx 9\%v$). The agreement is striking as, except for the slight difference in luminosity, one could think that the two pictures are parts of a bigger one. For comparison, the structure associated with a different $\rho_{D3}$-value is significantly different, as illustrated in the third picture.

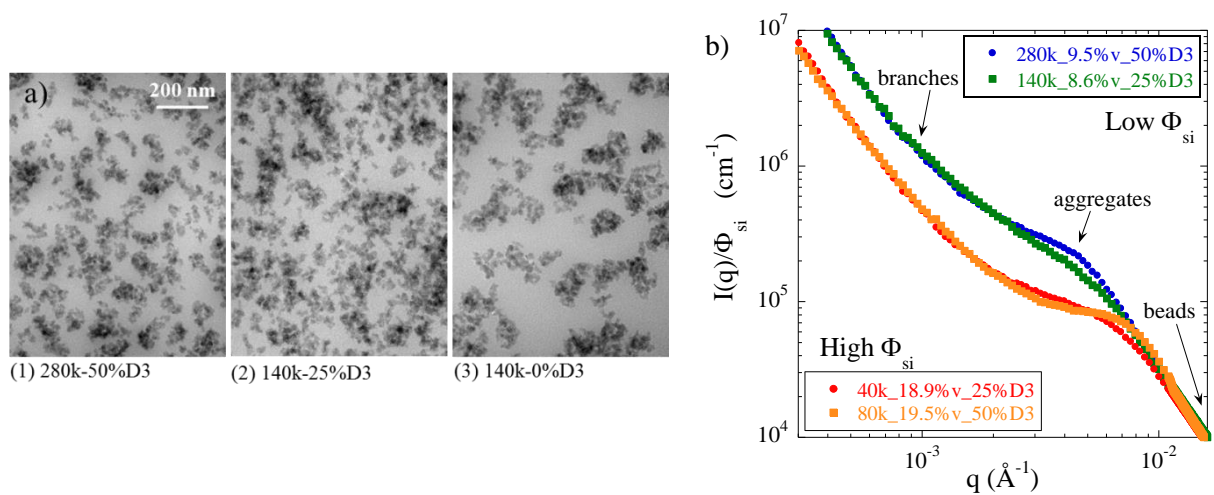

**Figure 4: (a)** TEM of nanocomposites with $\Phi_{si} \approx 9\%v$. Pictures **(1)** and **(2)** are for twin-samples with $\rho_{D3} = 35\times10^{-3}$ nm$^{-2}$. For comparison, picture **(3)** shows a composite with different $\rho_{D3} = 0$ nm$^{-2}$. **(b)** Reduced SAXS scattered intensity for twin nanocomposites with low $\Phi_{si}$ ($\rho_{D3} = 35\times10^{-3}$ nm$^{-2}$), and high $\Phi_{si}$ ($\rho_{D3} = 51\times10^{-3}$ nm$^{-2}$). Arrows indicate the zones corresponding to different structural levels. Adapted with permission from G. P. Baeza et al. ACS Macro Letters 3 (2014) 448-452. Copyright (2014) American Chemical Society.

The SAXS intensities of the same low-$\Phi_{si}$ twin samples ($35\times10^{-3}$ nm$^{-2}$), as well as a second couple ($51\times10^{-3}$ nm$^{-2}$) with higher silica fraction, are confronted in Figure 4b. In both cases, the superposition of the intensities is remarkable over almost the entire q-range, suggesting a very similar filler structure on this scale. In ref. [41], a numerical estimation of the deviations between intensity curves has been proposed. For the different twin-samples, the intensity deviations are found considerably smaller for twins than for all other samples, confirming quantitatively that the resemblance of the scattering curves in Figure 4b is not only a visual effect. TEM and SAXS results thus evidence that twin samples, defined by different mass and grafting, but same $\rho_{D3}$, display an identical microstructure. It can be concluded that the grafting density is the structure-determining parameter in simplified industrial nanocomposites, and not the chain mass. [41] Note that this agrees nicely with current knowledge in model systems, where the effect of grafting density on the composite morphology has also been identified. [47]

One of the fundamental questions of nanocomposites is to understand and control the influence of the structure on the mechanical properties. We have used dynamical mechanical analysis (DMA) to measure the storage and loss moduli of twins. As illustrated in the inset of Figure 5, the curves have the classical appearance of moduli, with a maximum due to the segmental relaxation in G''(T) and a viscoelastic (rubbery) plateau above the glass transition in G'(T). The plateau modulus values at fixed temperature and frequency are reported in Figure 5 for many twin-samples.

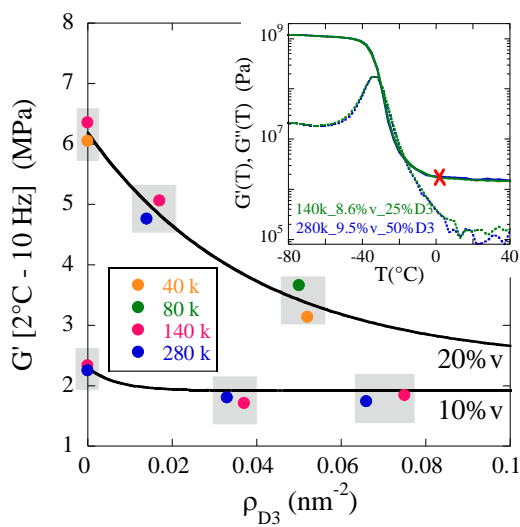

**Figure 5:** Viscoelastic plateau moduli at 2°C/10 Hz from DMA for twin samples. The corresponding chain masses are indicated in the graph. Shadowed areas identify twins. Inset: DMA curves (G' solid, G'' dotted line) for a twin couple with $\rho_{D3}$ = 35×10$^{-3}$ nm$^{-2}$. Reproduced with permission from G. P. Baeza et al. ACS Macro Letters 3 (2014) 448-452. Copyright (2014) American Chemical Society.

Clearly, the storage moduli were identical within 10% for each couple of twins. At 10%v of silica, they were close for all samples, whereas at 20%v a strong decrease was observed with increasing $\rho_{D3}$, proving the tunability of the mechanical performance (reinforcement) at high filler fraction. Twins have thus an identical structure and viscoelastic plateau modulus, but display different dynamics in the terminal regime due to the widely different chain masses. [41] This suggests that it is possible to control the onset of flow behaviour of twin samples over a wide frequency range by modifying the chain mass at fixed $\rho_{D3}$.

### III.5 A general scheme of aggregate formation during mixing

In silica of industrial origin, nanoparticles are aggregated, and aggregates are agglomerated in pellets. Depending on the torque during the mixing, these agglomerates are destroyed, aggregates are possibly broken into pieces, and these objects may re-aggregate due to H-bonds and Van der Waals interactions in the melt. A generalized model view of this process based on the density of grafted polymer mediating steric particle interactions is presented here, in agreement with the existence of twin samples discussed above. [40] Instead of comparing the full SAXS intensity curves, the average aggregate radii and aggregation numbers obtained with the structural model (section III.1) have been compared. These data are plotted in Figure 6 as a function of $\rho_{D3}$, in two subsets, for 9.5%v and 19%v of silica.

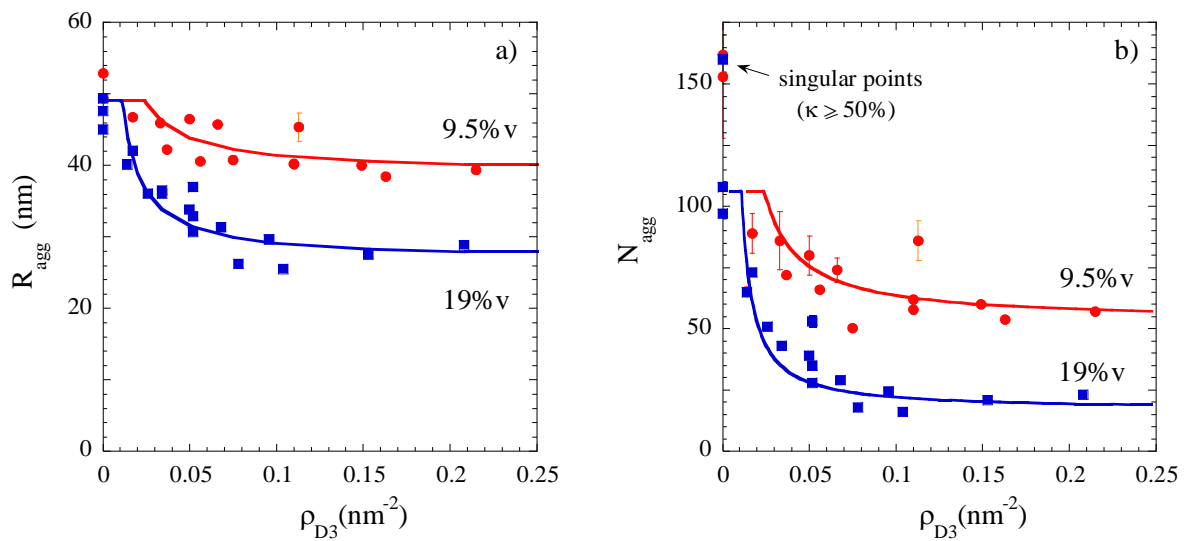

**Figure 6:** Average aggregate radius **(a)**, and average aggregation number **(b)** as a function of the nominal grafting density for composites with $\Phi_{si} \approx$ 9.5%v (red) and 19%v (blue). Data are compared to the model prediction (lines) using $\kappa$ = 35%. Error bars are related to scattering in the fraction of branches determined by TEM. Adapted from Ref [40] with permission from the Royal Society of Chemistry.

The data points in Figure 6 are rather scattered for both data sets, but a general tendency can be recognized. Both $R_{agg}$ and $N_{agg}$ functions start from about 50 nm and 100, respectively, then decrease, and level off for higher grafting densities. Such a decrease followed by saturation has already been observed in the study of the grafting fraction in section III.2. In addition, the higher silica fractions lead to smaller and lighter aggregates, at least at high $\rho_{D3}$, presumably due to a better break-down of aggregates at high viscosity in the mixing process.

The physical mechanism relating grafting density and structure formation in the mixer may be considered. $\rho_{D3}$ is a nominal value giving the number of grafts per unit surface on each (individual) silica NP. If NPs are aggregated, it is supposed that grafting will preferentially take place at the aggregate surface, defining a new aggregate grafting density, $\rho_{D3}^{agg}$. The latter depends on the aggregate structure, and is expected to be higher than the nominal NP grafting density due to the lower specific surface. In the mixing process, initial aggregates with radius $R_{max}$ are broken into smaller ones of typical radius $R_{min}$, depending on the maximum torque in the mixer. On these aggregates, some grafting takes place, and depending on the resulting aggregate grafting density, aggregation can take place or not. Once the aggregate grafting density reaches a critical aggregate grafting density $\rho_{D3}^c$, the re-aggregation process is stopped. It follows that aggregate radii are set by the aggregate grafting density reaching a limiting value, $\rho_{D3}^{agg} = \rho_{D3}^c$, via the establishment of a grafted brush regime. We have constructed a simple model [40] to rationalize the evolution of the aggregate size:

$$R_{agg}(\rho_{D3}) = \begin{cases} R_{max} \text{ for } \rho_{D3} < \rho_{D3}^m \\ R_{min} + \dfrac{R_{si}}{\kappa \frac{\rho_{D3}}{\rho_{D3}^c}} \text{ elsewhere} \end{cases} \quad (2)$$

$R_{si}$ is a radius representative of the silica bead (here $R_{si} = R_0$), and $\rho_{D3}^m$ is a cut-off value in grafting density below which $R_{agg}$ has been set to $R_{max} = 49.2$ nm, for both $\Phi_{si}$. As the compacity stayed approximately constant for the different samples, we used a unique $\kappa$ value of 35% for simplicity. The results are shown in Figure 6, with fair agreement with the data. A critical grafting density of $\rho_{D3}^c$ = 0.01 nm$^{-2}$ was found, which is quite close to $1/R_g^2$ = 0.007 nm$^{-2}$ for $M_{SB}$ = 140 kg.mol$^{-1}$, the de Gennes estimate of the onset of the mushroom-to-brush transition [48, 49].

### III.6 Independent cross-check of structure by swelling of nanocomposites

In order to check the structural model (section III.1) describing interacting aggregates, it would be desirable to produce dilute nanocomposite samples in order to focus on individual non-interacting aggregates. Unfortunately, direct dilution is impossible because of large differences in viscoelasticity between the composite and the pure polymer, and solid mixing of both components generates only a dispersion of the original material in the polymer matrix. Therefore, an original method to dilute the filler to low-volume fractions by swelling of nanocomposites has been developed.

Swelling nanocomposites with solvents or monomers has been used in the field of intercalating nanoclays [50-52], but also in SB nanocomposites with industrial filler in organic solvents to draw qualitative conclusions on the absence of interactions between aggregates [36]. Here, we combined matrix crosslinking, swelling in a good monomer solvent (styrene), and post-polymerization of these monomers, to demonstrate that it is possible to separate the filler into small aggregates. [43]

The complete swelling protocol is detailed in ref [43]. The silica fraction in the final swollen samples was about 1.5%v starting from nanocomposites with $\Phi_{si}$ in the range 8.2 – 21.3%v (swelling ratio around 10). The silica microstructure of these samples has been studied by scanning electron microscopy (SEM) and SAXS as shown in Figure 7.

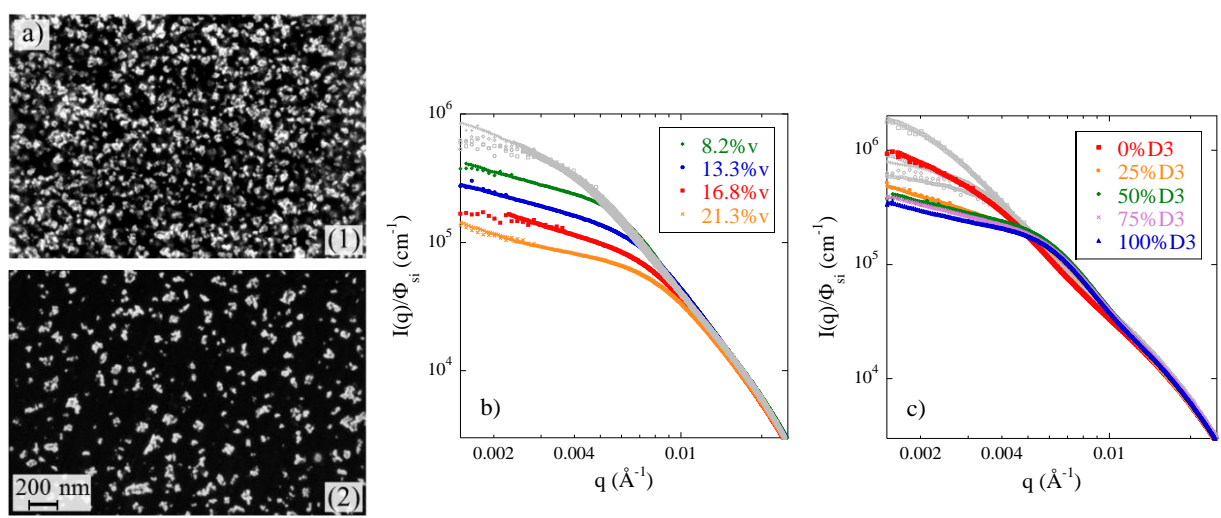

**Figure 7: (a)** SEM of a nanocomposite with $\Phi_{si}$ =16.8%v (50%D3, 190 kg.mol$^{-1}$) **(1)**, and of the same sample after swelling ($\Phi_{si}$ = 2.0%v) **(2)**. **(b)** Reduced SAXS scattered intensity for unswollen (colored symbols) and swollen (grey symbols) nanocomposites for the series in silica fraction at 50%D3 (190k). The fractal network contribution has been subtracted. **(b)** Same data for the series in matrix composition at 8.2%v silica (190k). Adapted from Ref [43] with permission from the Royal Society of Chemistry.

The SEM pictures of both the original (1) and swollen (2) nanocomposites are compared in Figure 7a. The homogeneous network-type structure in picture (1) is broken up on the scale of aggregates, which appear to be well-dispersed in space upon swelling. Aggregates stay in the same size range of typically below 100 nm, with a clear polydispersity in size and shape. In Figure 7b, two families of normalized SAXS curves – unswollen and swollen series with various original $\Phi_{si}$ values – are plotted after subtraction of the low-q network contribution, in order to apply the structural model in the intermediate-q range. In both cases, the intensities are seen to level off to a quasi-plateau at intermediate q. The impact of the silica content in nanocomposites has been summarized in section III.1, and we find again a clear decrease of the plateau level with increasing $\Phi_{si}$. On the opposite, the intensities of swollen samples overlap rather nicely over almost the entire q range. It demonstrates that diluting the sample via the swelling procedure amounts to reducing interactions and shows that aggregates are mostly identical in spite of the different filler fractions of the initial formulation. The situation is completely different for the series with various matrix compositions at fixed silica fraction. The corresponding SAXS curves are shown in Figure 7c, before and after swelling. As in section III.2, the decrease in intensity for the original samples is due to the reduction in aggregate mass. Data obtained in Figure 7c for the swollen samples confirm the change in aggregate mass. In both cases ($\Phi_{si}$ and %D3 variations), the results obtained on swollen nanocomposites thus validates the tendencies observed with the original structural model applied directly to the highly loaded cases. [43]

Ideally, swelling was expected to reduce the inter-aggregate structure factor to one (diluted case), and to allow measuring the aggregate form factor. It turns out, however, that the very low-q scattering does not overlap perfectly in Figure 7b. The aggregate size determination is hindered by the remaining interferences of the structure factor for the swollen samples. In such a case of intermediate dilution, we must conclude that even if the qualitative result in Figure 7b is convincing, the structural analysis is not appropriate for extracting compacity and aggregate characteristics from these scattering curves of swollen samples with high precision.

### III.7 Limitations and possible improvements of the structural model

The original structural model provides a quantitative interpretation in terms of aggregate size and mass of the multiscale structure based on interacting aggregates making up large-scale fractal branches in industrial nanocomposites [38]. Despite its robustness, we are nonetheless aware of several limitations, which are discussed point by point below.

The first point deals with the (unknown) nature of the interaction between aggregates. In the model, we have considered the simplest one, i.e., excluded volume (hard spheres) interaction [53, 54]. However, due to their spread-out structure (as in Figure 2a), aggregates may to some extent interpenetrate, and one could rather consider a softer repulsive interactions [55]. Application of our model using a different potential would of course lead to different values of, e.g., aggregate compacity (softer potentials giving lower masses), but without altering significantly the relative evolutions from one sample to the other. Secondly, even if the choice of the (a priori unknown) aggregate polydispersity in size seems reasonable, small variations in $N_{agg}$ have to be considered with caution. One should also note that an evolution of the aggregate size distribution toward higher monodispersity may be envisaged for the series in matrix composition, as one might guess from the TEM pictures in Figure 2a and from the scattering data in Figure 2b. A lower polydispersity could lead to higher compacities, which would reduce the variation of κ with %D3. Finally, due to the limitation of the simulations, it is not possible to simulate a system with real aggregate fractions. For this reason, we could only use the extrapolation of the low-q limits of the structure factors calculated for non-concentrated systems ($\Phi_{agg}$ below 30%).

## IV. Dynamical processes in simplified industrial nanocomposites

Beyond structure, dynamical properties of nanocomposites are of equal importance for applications like car tire optimization. The viscoelastic response of nanocomposites is strongly temperature- and frequency-dependent, which is exploited in modern car tires in order to decrease (low frequency-) rolling resistance – and thus fuel consumption – without sacrificing (high frequency-) wet grip. In this section, we review results of broadband dielectric spectroscopy (BDS) to characterize in detail the dynamical processes in nanocomposites with multi-scale filler structure. The latter processes include ionic conductivity as well as a high-temperature interfacial process, which are both found to be impacted by structural reorganizations, in close relationship with the mechanical properties [24, 42]. Such results make obvious that microstructure, i.e., the dispersion state of the filler, sets the frame for the dynamical properties. In this context, the question of the immobilized polymer layers as measured by NMR has also been considered [42].

### IV.1 A new dielectric interfacial process as a probe of large-scale filler structure

The dielectric response in the high-T range (well above the glass transition) is illustrated in Figure 8 for two series of nanocomposites with different silica fractions and different matrix compositions. In all cases, one can clearly see in the dielectric loss $\varepsilon''(\omega)$ a low-frequency upturn characteristic of the ionic conductivity, and a broad maximum where two distinct slow dielectric processes called Maxwell-Wagner-Sillars processes have been identified. MWS processes are related to polarization effects due to charge carrier diffusion through the different phases of nanocomposites. Such a polarization results in the trapping and accumulation of charges at the polymer/silica interface, which is why they do not exist in the pure matrix (inset in Figure 8a). As silica NPs are non-conducting, ions can diffuse on the surface of individual NPs within aggregates, or on a larger scale, either on the aggregate surface, or across polymer between aggregates. For slow, large-scale processes, the corresponding time scales are controlled both by the diffusion rate of charges and the characteristic distances to be covered, the latter being given by the multi-scale filler structure. The aim of the study was to correlate the evolution of the dynamical dielectric properties of these interfacial processes with the different scales of the microstructure. Such an approach is original in the sense that dielectric spectroscopy is obviously not a spatially resolved technique, and it is therefore rarely [56] used to provide structural information.

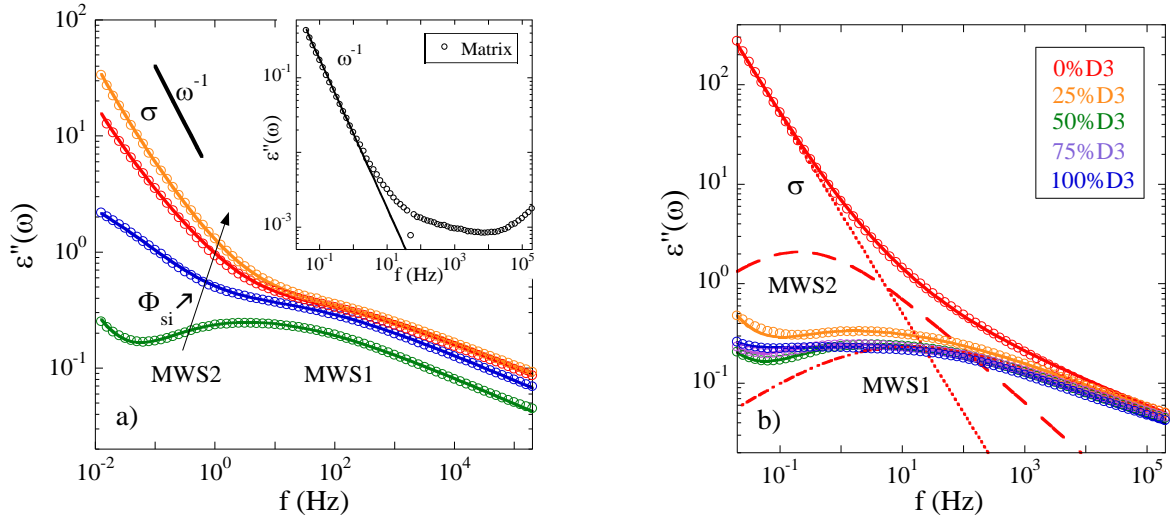

**Figure 8:** Dielectric loss spectra at 333 K for nanocomposites with different $\Phi_{si}$ (50%D3, 140 kg.mol$^{-1}$) **(a)**, and different amounts of graftable chains ($\Phi_{si} \approx 8.5\%v$, 140 kg.mol$^{-1}$) **(b)**. Solid lines are fits by the sum of two Havriliak-Negami functions and a dc-conductivity term. Individual contributions are included for the sample with 0%D3. Inset: same data for the pure matrix. Reproduced from Ref [24] with permission from the PCCP Owner Societies, and from Ref [42] with permission from Polymer.

The low-T interfacial process (MWS1) – on the high-frequency side in Figure 8 – has been previously observed in several composite systems containing carbon black or silica [56-59]. With silica, an interlayer model [60] based on the diffusion of charges in hydration layers surrounding silica NPs has been applied to industrial rubber-based nanocomposites [57]. Such a model could also be used to describe our data, indicating that MWS1 relates to a rather local phenomenon on the scale of a few NPs. The second high-T interfacial process (MWS2) – on the low-frequency side in Figure 8 – is a new structure-related process, which has been evidenced for the first time in silica-filled nanocomposites [24]. Note that it has been partially identified for carbon black and related to reorganizations upon heat treatment and vulcanization. [59] The MWS2 process has been characterized following the fitting procedure described in ref [24], and we found activation energies independent of the silica fraction ($E_{MWS2}$ = 114 kJ.mol$^{-1}$ on average), and comparable to that associated with the temperature dependence of the ionic conductivity. This similarity suggests that the dynamics of MWS2 is primarily controlled by the polymer matrix, as found for the ionic conductivity. On the other hand, both the conductivity and the MWS2 time scales (not reported here) depend on the silica fraction, indicating that the dynamics of MWS2 also relates to the nanocomposite structure. There are thus two different conduction mechanisms, one through the matrix between aggregates, and another one along the surface of aggregates. Turning to the dielectric strengths $\Delta\varepsilon_{MWS}$, both interfacial processes display temperature-independent values. The average values are reported in Figure 9 as a function of filler fraction and matrix composition, together with the ionic conductivity obtained at the highest temperature available.

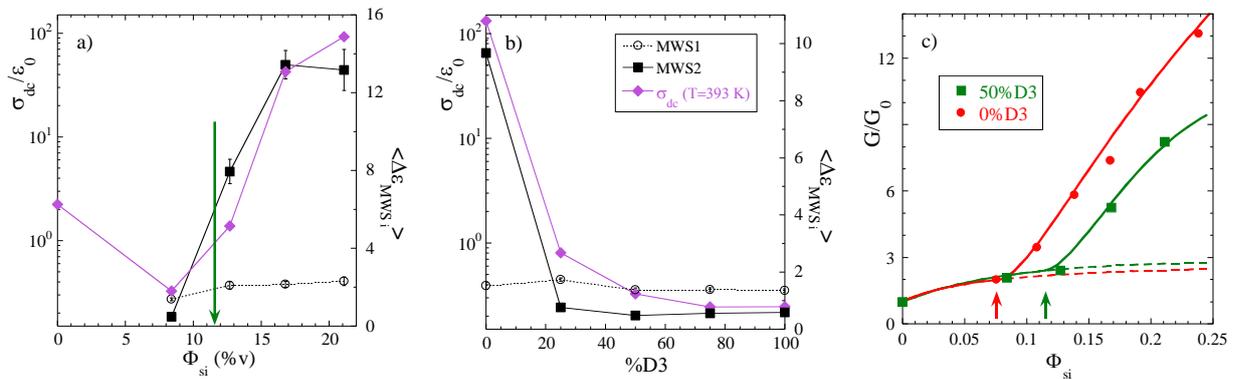

**Figure 9:** DC-conductivity at 393 K (diamonds) and average value of the dielectric strengths over T for MWS1 (circles) and MWS2 (squares) as a function of filler fraction (50%D3, 140 kg.mol$^{-1}$) **(a)**, and matrix composition ($\Phi_{si} \approx 8.5\%v$, 140 kg.mol$^{-1}$) **(b)**. Lines are a guide to the eye. The percolation threshold as obtained from rheology is indicated by an arrow. **(c)** Rheological response at 60 Hz in terms of the reinforcement factor $G/G_0$ as a function of silica fraction for 140k-nanocomposites with 0%D3 (circles) and 50%D3 (squares). Solid lines are fits discussed in the text, arrows indicate the percolation threshold. The purely hydrodynamic reinforcement, $1+2.5\Phi_{agg}$, is also included (dotted lines). Data are taken from refs [37, 38]. Reproduced from Ref [24] with permission from the PCCP Owner Societies, and from Ref [42] with permission from Polymer.

Figure 9 shows that whereas the MWS1 process is rather insensitive to the silica content and the fraction of graftable chains, both the ionic conductivity and MWS2 display a strong evolution with these parameters. The first decrease of $\sigma_{dc}$ between 0 and 8.4%v in Figure 9a is compatible with the trapping of free charge carriers initially present in neat polymer at the silica surface. Above 8.4%v, however, the contribution of free charge carriers cannot be invoked, and the strong conductivity increase (by more than two orders of magnitude) points towards the formation of large clusters or of a percolated network of aggregates through the sample, as already reported by Medalia some 30 years ago [61]. Interestingly, an opposite behaviour is observed in Figure 9b, where the conductivity strongly decreased with %D3. This transition with grafting is interpreted as the depercolation of aggregates. The BDS response in Figure 9 shows that the new interfacial process MWS2 is correlated with these global structural features. Also, MWS2 and MWS1 processes are associated with a broad distribution of relaxation times (Figure 8b), in agreement with such structurally polydisperse systems.

Percolation of the filler structure has direct consequences on the rheological response to shear. One of the characteristics of the G'(ω) curves is the high-frequency plateau modulus G, which can also be determined for the unfilled matrix ($G_0$), allowing to define the reinforcement factor as the ratio of moduli $G/G_0$. In Figure 9c, the evolution of this reinforcement factor with silica fraction is compared for two matrix compositions, 0%D3 and 50%D3. To describe these data, a percolation model [62, 63] has been adapted introducing percolation of aggregates inside the branches, which themselves extend across the whole sample (see ref [38] for details). In this picture, the aggregate fraction $\Phi_{agg}$ within the branch becomes the main variable, instead of the silica fraction. $\Phi_{agg}$ values were calculated using the aggregate compacity and the fraction of fractal branches as determined by TEM and SAXS. Given the good quality of the data description in Figure 9c, the compatibility with our structural analysis by SAXS underlines the consistency of our method.

The critical percolation fraction of aggregates in the branches was found to be $\Phi_{agg}^c$ = 56%v at 50%D3, which is consistent with our picture of aggregates percolating within the fractal branches, i.e., in a space of reduced dimension. It corresponds to an overall silica fraction of $\Phi_{si}^c$ = 11.5%v, in good agreement with the strong conductivity increase observed between 8.4 and 16.8%v in Figure 9a, and thus also with the increase of dielectric strength for MWS2. Nanocomposites without graftable chains (0%D3) displayed a lower percolation fraction, $\Phi_{si}^c$ = 7.5%v. As already evidenced by the visual inspection in Figure 9c, polymer grafting moves the system away from percolation by inducing a better dispersion of the filler. Such a result agrees nicely with literature results, see e.g. [5] for a recent review of the different regimes arising as a function of the grafted chain mass and grafting density with respect to (short or long) matrix chains, and their impact on the dispersion state. The shift of the percolation threshold observed here is also in qualitative agreement with the strong conductivity decrease – and decrease of $\Delta\varepsilon_{MWS2}$ – observed between 0 and 25 %D3 in Figure 9b.

Evidence has thus been provided that the new high-T dielectric process MWS2 is a sensitive tool to large-scale structural reorganizations, and in particular percolation of the aggregates of silica NPs within space-filling branches. More specifically, the strength of MWS2 follows the percolation behavior observed both in rheology and in the ionic conductivity, whereas the MWS2 activation energy hints at a control of the dynamics by diffusion processes through the matrix between filler aggregates.

## IV.2 Immobilized polymer layers

The existence of polymer layers between aggregates is related to the question of immobilized, often referred to as "glassy", layers which are mainly detected by NMR [64]. Such a topic relates to the local segmental dynamics, and it has been intensely studied by several groups in the past years [45, 57, 65-67]. However, these dynamical properties have only rarely been compared to macroscopic rheological properties [23, 56, 59, 67]. And so, it remains unclear if the reinforcement of mechanical properties is dominated by local dynamical heterogeneities bridging NPs [68-70] or large-scale filler structure [71].

In our nanocomposites, the segmental ($\alpha$-) relaxation has been investigated by BDS, as a function of silica content up to 21.1%v and polymer grafting. We found that the spectral shape (shape parameters and the dielectric strength) of the $\alpha$-process remains unchanged as compared to the dielectric response of the pure matrix. As can be seen in Figure 10a, the time-scale for the segmental dynamics is also not modified with $\Phi_{si}$ and %D3 (see inset), which means that either there are no immobilized layers, or BDS is not sensitive enough, possibly due to superposition with MWS and conductive processes. Note that in systems with strong interactions, e.g. poly(vinyl acetate) adsorbed on silica, detection of a layer of reduced mobility by BDS was successful.[72] In other cases, surfactant acting as BDS probe increased the sensitivity to interfacial polymer dynamics.[73]

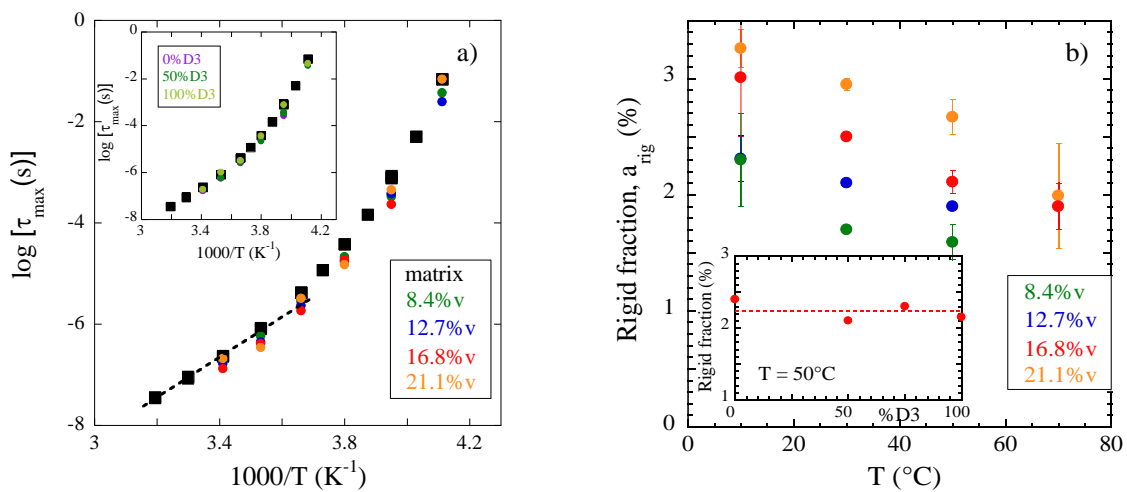

**Figure 10: (a)** Temperature dependence of the relaxation times for the $\alpha$-process in 140k-nanocomposites with $\Phi_{si}$ = 8.4 – 21.1%v and 50%D3. Inset: same data for different %D3 at fixed $\Phi_{si} \approx$ 8.5%v. **(b)** NMR-determined fraction of immobilized polymer as a function of temperature for the same nanocomposites. Inset: rigid fraction versus %D3 at 50°C and fixed $\Phi_{si} \approx$ 16.7%v. Reproduced from Ref [42] with permission from Polymer.

To check the existence of an immobilized fraction of organic material in our nanocomposites, we have performed low-field proton NMR experiments. Note that the characteristic time for segmental dynamics as seen by NMR stands in the µs range. Thus, the temperature range investigated is just above the "NMR $T_g$", i.e., about 40 K above the calorimetric $T_g$ [64]. The apparent amount of rigid polymer is plotted in Figure 10b as a function of temperature. It was found to increase with $\Phi_{si}$ and decreased with temperature as expected due to the softening of the immobilized part. The obtained fractions were rather low (between ≈ 1.5 and 3.5%) as compared with other systems like e.g., silica-polyethylacrylate nanocomposites [45], but in good agreement with a recent work by Mujtaba et al. [68] on industrial SB rubber nanocomposites. We can rationalize these results based on the structure of aggregates filling fractal branches in our system, and calculate the "glassy" polymer fraction in an aggregate. At 30°C, a $\Phi_{si}$-independent value of (7.3 ± 0.7)% was obtained. A simple estimation of the layer thickness based on the hypothesis of individual bead dispersion gave very thin immobilized layers of about 0.3 nm. This remains also true in presence of small aggregates, like here. Finally, there was no detectable variation of the immobilized fraction as a function of the quantity of graftable chains (see the inset in Figure 10b), presumably due to the too low grafting densities. The low fraction of immobilized polymer obtained from NMR confirms the BDS results, where it has been found that the α-relaxation was not affected by the presence of silica in our concentration range [42].

Incidentally, one may also comment on the NMR results in the light of the reinforcement factors shown in Figure 9c. It has been shown recently by Mujtaba et al. [68] that reinforcement may be coupled with the fraction of immobilized polymer layers on the surface of the NPs. The interpretation is that even thin layers convey connections between the NPs. In the present case, the fraction of (equally thin) immobilized layers is found to be independent of the grafting fraction. On the other hand, the rheological data shown in Figure 9c correlate with the strong structural and dynamical evolution as seen by BDS, both with silica content and grafting. It may thus be concluded that the structural contribution – and not the quantity of immobilized polymer – dominates the rheological response in our samples [42].

## V. Conclusion

A coherent body of work recently contributing to the fundamental understanding of the multi-scale structure and dynamics of silica-SB simplified industrial nanocomposites has been reviewed. The use of combinations of complementary techniques, TEM and small-angle scattering for structure, and rheology, NMR and BDS for dynamics, gives insights which would not have been achievable with a single technique: percolation, e.g., is a structural property, but scattering alone is unable to decide on this issue, whereas it was clearly identified in the BDS-study. "Simplified" refers to the absence of any cross-linker and related catalysts. By construction, these systems thus bear a certain model character, but are otherwise close to applications due to the use of a highly disordered industrial filler and a solid-phase mixing protocol. An original *quantitative* approach based on a global SAXS/TEM model has been used in order to describe the structural data. It is found that primary silica beads (typical size 17 nm) are organized within small aggregates (typical size 80 nm, aggregation number about 50, giving an internal volume fraction of the order of 35%), which themselves fill large-scale fractal branches (typical width 150 nm). The evolution of aggregate structure upon the variation of all experimental parameters (chain mass, fraction of graftable chains, silica content) led to the main finding that the size and density of filler aggregates may be tuned by the grafting density of SB chains, whereas these characteristics remain largely insensitive to filler fraction in spite of the mechanical mixing process used to formulate such samples. In this context, conclusions on the mechanisms of aggregate formation could be inferred: once the grafting density exceeds a critical value, even the smallest possible aggregates are covered by a polymer brush preventing re-agglomeration during the mixing process. The reliability of these structural results has been cross-checked with an independent method based on swelling of the nanocomposites together with SAXS. With the aim of connecting structure and dynamical properties, simplified industrial nanocomposite samples have been further characterized by dielectric spectroscopy and NMR. A new high-temperature dielectric process MWS2 has been evidenced. Knowledge of the microstructure and linear rheology of these nanocomposites allows to correlate its evolution with the structure. Similar to rheology, the MWS2-process is sensitive to depercolation upon polymer grafting. This establishes that dielectric spectroscopy can be used as a simultaneous probe for large-scale structural reorganizations in complex silica-polymer nanocomposites and local dynamics. Finally, it has been shown that dynamical heterogeneities ("glassy" layers) have a minor impact on the mechanical response as compared to structural contributions in our samples.

## Acknowledgements

The authors are thankful for support by ANR under Contract No. ANR-14-CE22-0001-01 (NANODYN). Fruitful discussions with M. Couty (Michelin) are gratefully acknowledged. Finally, progress in the field of polymer nanocomposites benefits greatly from small-angle scattering beamtimes granted by large-scale facilities, in particular ESRF (Grenoble), Soleil (St Aubin), LLB (Saclay), ILL (Grenoble), and MLZ (München).

Graphical Abstract

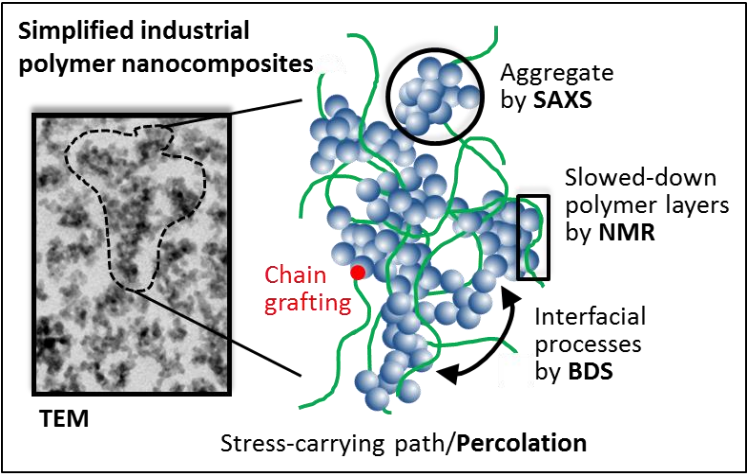